\documentclass[trackchanges]{aastex7}

\usepackage{tabularx}
\usepackage{longtable}
\usepackage{multirow}
\usepackage{booktabs}
\usepackage{amsmath}

\definecolor{DarkGreen}{rgb}{0.0,0.4,0.0}  
\newcommand{\B}[1]{{\color{blue}{#1}}}
\newcommand{\R}[1]{{\color{red}{#1}}}
\newcommand{\G}[1]{{\color{DarkGreen}{#1}}}
\newcommand{\M}[1]{{\color{magenta}{#1}}}
\newcommand{\C}[1]{{\color{cyan}{#1}}}
\usepackage[normalem]{ulem}
\usepackage{soul}
\usepackage{cancel}

\begin{document}

\title{Investigation on the Predictive Potentiality of Photospheric Magnetic Parameters for Distinguishing Confined from Eruptive Solar Flares}

\author[orcid=0009-0008-3359-8092, gname=Jinhui, sname='Pan']{Jinhui Pan}
\affiliation{CAS Key Laboratory of Geospace Environment, Department of Geophysics and Planetary Sciences, University of Science and Technology of China, Hefei 230026, People’s Republic of China}
\email{jhpan@mail.ustc.edu.cn}  

\author[orcid=0000-0003-4618-4979, gname=Rui, sname='Liu']{Rui Liu} 
\affiliation{CAS Key Laboratory of Geospace Environment, Department of Geophysics and Planetary Sciences, University of Science and Technology of China, Hefei 230026, People’s Republic of China}
\affiliation{Mengcheng National Geophysical Observatory, University of Science and Technology of China, Hefei 230026, People’s Republic of China}
\email[show]{rliu@ustc.edu.cn}

\begin{abstract}

A question often arises as to why some solar flares are confined in the lower corona while others, termed eruptive flares, are associated with coronal mass ejections (CMEs). Here we intend to rank the importance of pre-flare magnetic parameters of active regions in their potentiality to predict whether an imminent flare will be eruptive or confined. We compiled a dataset comprising 277 solar flares of GOES-class M1.0 and above, taking place within 45 deg from the disk center between 2010 and 2023, involving 94 active regions. Among the 277 flares, 135 are confined and 142 are eruptive. Our statistical analysis reveals that the magnetic parameters that are most relevant to the flare category are: total unsigned magnetic flux $\Phi$, mean magnetic shear angle $\Theta$ along the polarity inversion line (PIL), photospheric free magnetic energy $E_f$, and centroid distance $d$ between opposite polarities. These four parameters are not independent of each other, but in combination, might be promising in distinguishing confined from eruptive flares. For a subset of 77 flares with high-gradient PILs, the area of high free energy regions ($A_{\mathrm{Hi}}$) becomes the most effective parameter related to the flare type, with confined flares possessing larger $A_{\mathrm{Hi}}$ than eruptive ones. Our results corroborate the general concept that the eruptive behavior of solar flares is regulated by an interplay between the constraining overlying flux, which is often dominant in both $\Phi$ and $E_f$ and related to $d$, and the current-carrying core flux, which is related to $\Theta$.

\end{abstract}



\section{Introduction\label{S-Introduction}}
Solar flares and coronal mass ejections (CMEs) are the most energetic phenomena in the solar system. Solar flares are usually classified as eruptive and confined types according to their association with CMEs \cite[]{2001ApJ...552..833M, 2007ApJ...665.1428W}. CMEs mainly originate from the sheared and twisted magnetic field in the corona through magnetic reconnection  \cite[]{2000JGR...10523153F, 2011LRSP....8....6S}, propagate in interplanetary space, and pose a negative impact on the planetary environment \cite[]{Temmer2021}.

Numerous studies have investigated the relationship between the magnetic properties of solar active regions (ARs) and their association with solar flares and CMEs, with a generally accepted notion that whether an eruption is successful or failed is determined by the equilibrium balance between the upward-directed magnetic force, mostly exerted by the core eruptive structure, and the downward-directed magnetic force, mostly provided by the large-scale overlying field. \cite{2007ApJ...655L.117S} conducted an analysis of the magnetic properties of solar ARs associated with 2500 flares, including 289 M- and X-class flares, and found that large magnetic field gradients along the polarity inversion lines (PILs) as well as the total unsigned magnetic flux within 15 Mm of the PILs serve as reliable predictors for the occurrence of solar flares. \cite{Wang&Zhang2007ApJ...665.1428W} implemented a comparative study of four X-class confined flares and four X-class eruptive flares and concluded that the sites of confined flares are located closer to the weighted center of the magnetic flux distribution within the ARs, whereas eruptive flares tend to occur farther from this center. \cite{Cheng2011ApJ...732...87C} comparatively analyzed nine M-class and X-class flares in AR 10720, examining flare duration, displacement from the AR center, and the decay index of the overlying transverse field. They showed that six impulsive, centrally-located events remained confined, whereas three long-duration outskirts events triggered fast halo CMEs. \cite{DeRosa&Barnes2018ApJ...861..131D} found that X-class flares that don't have access to open magnetic fields are less likely eruptive. Based on the critical conditions of double arc instability \cite[DAI;][]{2017ApJ...843..101I}, \cite{2020Sci...369..587K} developed the so-called $k$-scheme method to evaluate the critical radius ($r_c$) of the trigger-reconnection region and the releasable free magnetic energy ($E_r$) within the high free energy regions (HiFERs) of the eruptive core regions. This method enables the prediction of both the location and timing of large-scale solar flare events. 

The launch of the Solar Dynamics Observatory \cite[SDO;][]{Pesnell2012} in 2010 has made possible the regular measurement of the photospheric vector magnetic field in high detail through the Helioseismic and Magnetic Imager \cite[HMI;][]{2012SoPh..275..207S}. Using a sample of 15 flare-active and 15 flare-quiet ARs, \cite{2020ApJ...893..123A} investigated the relationship between the flare/CME production of ARs and the electric current neutralization and magnetic shear along the polarity inversion line (PIL), and found that flare/CME-active ARs are less current neutralized and possessed a larger value of magnetic shear. The two flare types seem to be better discriminated by the degree of electric current neutralization, when the parameter is considered only in regions of high squashing factor \cite[]{2024ApJ...961..148L} or of selected magnetic flux structures based on their connectivities in nonlinear force-free field models \cite[]{2025ApJ...983L..28M}. These results highlight the informativeness of diagnostics focused on the core eruption structure. 

An important factor associated with the eruptive behavior of flares is the constraining effect of the magnetic field overlying the eruptive structure that often takes the form of a magnetic flux rope. The overlying field is also known as the magnetic cage \cite[]{2018Natur.554..211A} or strapping field \cite[]{Myers2015}. A magnetic flux rope is subject to the torus instability when the decay index ($n=-\partial \ln B_\mathrm{p} / \partial \ln z$) for the background field $B_{\mathrm{p}}$ overlying the rope reaches the critical value $n_{c}$ at a critical height $h_c$ \cite[]{Torok&Kliem2005ApJ...630L..97T, 2006PhRvL..96y5002K}. Many observational studies support the hypothesis that an eruption might be failed when the background field is strong enough \cite[e.g.,][]{Wang&Zhang2007ApJ...665.1428W, Liu2008ApJ...679L.151L, Cheng2011ApJ...732...87C, Sun2015ApJ...804L..28S, Thalmann2015ApJ...801L..23T, WangD2017ApJ...843L...9W, Toriumi2017ApJ...834...56T, Sun2022MNRAS.509.5075S}, but diverge on the parameters quantifying how strongly the background field confines the core structure. For example, \cite{WangD2017ApJ...843L...9W} found that the distance between the flux weighted center of the AR is highly correlated with the critical height through analyzing 60 flares of M-class and above. Based on database RibbonDB \cite[]{2017ApJ...845...49K}, \cite{2020ApJ...900..128L} showed that the total unsigned AR flux is an indicator of the strength of magnetic confinement: it anti-correlates with the portion of eruptive flares for a fixed GOES class,  and correlates positively well with the critical height. \cite{2021ApJ...917L..29L} further analyzed a large sample of 719 $\ge\,$C5.0 flares, of which 251 are eruptive and 468 confined, and found that confined flares display larger AR non-potentiality, including the length of steep-gradient PIL, photospheric free energy, and large magnetic shear area. All these non-potentiality parameters also correlate positively with the unsigned AR flux.  \cite{2022ApJ...926...56K} found that eruptive flares are associated with smaller PIL fluxes but larger magnetic shears than confined ones. 

As for flare properties, such as duration and magnitude, previous studies generally found that they are not pertinent to the eruptive behavior of a flare \cite[e.g.,][]{Harra2016SoPh..291.1761H,Toriumi2017ApJ...834...56T}, but flare ribbon properties might be promising. \cite{Toriumi2017ApJ...834...56T} investigated the flare duration, ribbon area, ribbon flux, etc., in 32 eruptive and 19 confined flares, and concluded that the ribbon-to-sunspot area ratio clearly determines whether a flare is eruptive or not, implying that CME productivity might be regulated by the interplay between the strapping field and the flaring source region. By analyzing 480 flares of GOES-class C5.0 and above on their X-ray thermodynamic properties and AR and flare-ribbon magnetic field properties, \cite{2023ApJ...958..104K} found that confined and eruptive events have similar reconnection fluxes but confined flares are more compact, occur in larger ARs, and show higher peak reconnection rates. Magnetic twist has also been considered as a property associated with the core eruptive structure, in the context of the helical kink instability \cite[]{Torok&Kliem2005ApJ...630L..97T}, but is often found to be not sufficient nor necessary for CME productivity \cite[]{Falconer2002ApJ...569.1016F,Bobra&Ilonidis2016ApJ...821..127B,Jing2018ApJ...864..138J,Zhou2019ApJ...877L..28Z}. In a study of magnetic flux ropes in 45 flares of GOES-class M3.9 and above, \cite{Duan2019ApJ...884...73D} argued that the kink instability plays an equally important role as the torus instability in regulating the flare eruptivity. Based on a study of 106 flares of GOES-class M1.0 and above, \cite{2022ApJ...926L..14L} proposed to use $\alpha/\Phi_\mathrm{AR}$ to measure the probability for a large flare to be eruptive or not, combining the magnetic twist parameter $\alpha$ within the flaring PIL region or within the region of high magnetic free energy density and the total unsigned flux of the AR, $\Phi_\mathrm{AR}$, which corresponds to the local overlying strapping field as opposed to the global one \cite[]{Sun2022MNRAS.509.5075S}. In a similar spirit, \cite{Lin2020ApJ...894...20L} proposed the ratio of the magnetic flux with twist higher than a threshold to the surrounding flux. \cite{2022SoPh..297...59K} reviewed SDO-era results on reconnection fluxes and rates, dimmings, and photospheric field changes, and argued that such flare-ribbon related and intensive parameters are relevant to the eruptive behaviors of flares. The problem with flare-ribbon related properties, however, is that they cannot be obtained until the event has begun, which limits their predictive ability. To predict whether the next flare within an AR will be more likely eruptive or confined, one has to resort to the pre-flare magnetic properties of the AR, without knowing which part of the AR would flare in the near future.


Thus, from the perspective of flare/CME forewarning and forecasting, we computed both AR-wide and PIL-focused photospheric parameters from pre-flare vector magnetograms, conducted a systematic statistical analysis to assess the significance of the relationship between these parameters and flare eruption types, and further quantified the strength of the association through non-parametric statistical tests. The rest of the paper is organized as follows: \S\ref{S-dm} describes the dataset and parameters we used; \S\ref{S-sre} outlines the statistical methods and presents the results; and \S\ref{S-discussion} provides a summary of the conclusions.

 
\section{Data and Method}\label{S-dm}      

\subsection{Data} \label{ss-d}
\begin{figure} [ht]
\centerline{\includegraphics[width=0.5\textwidth,clip=]{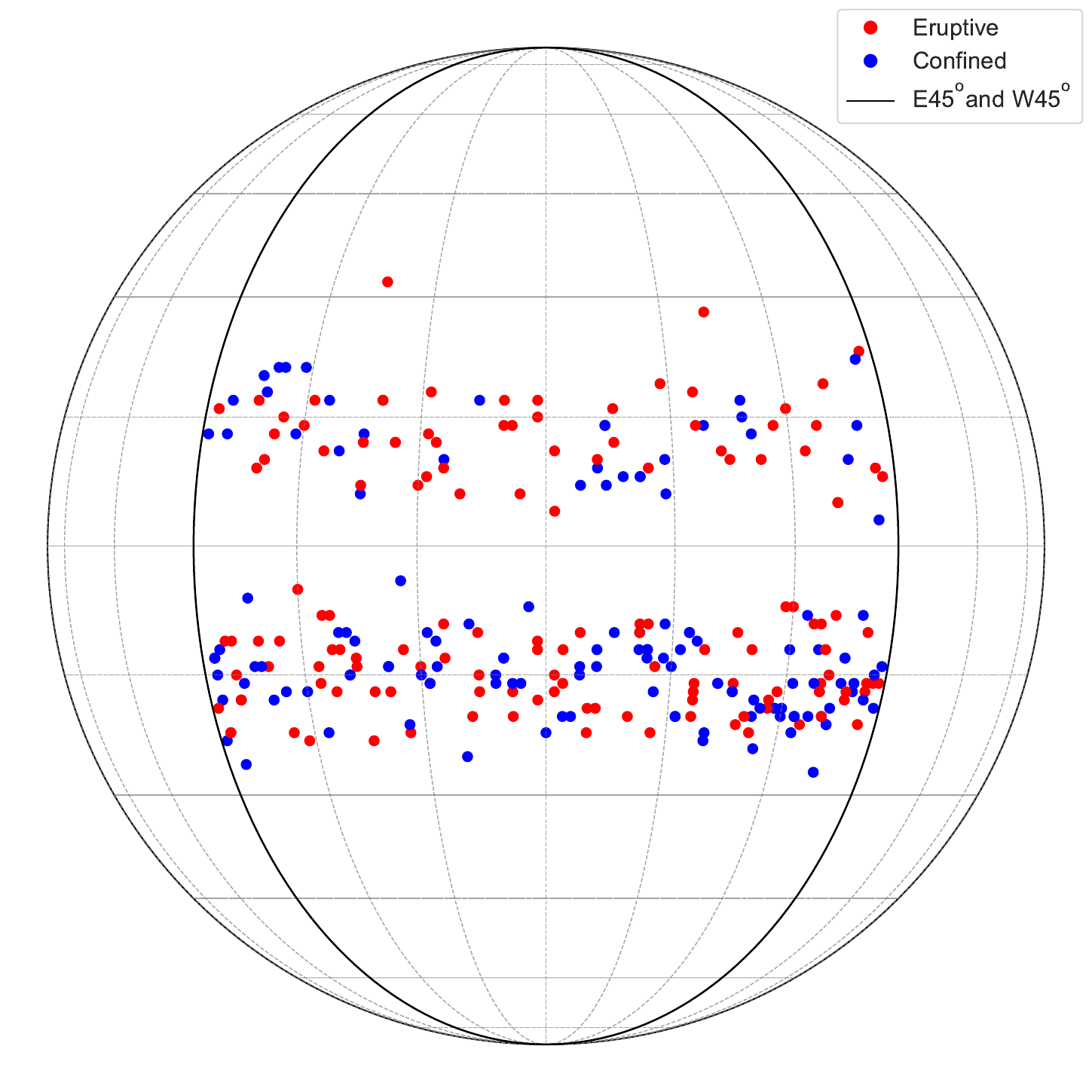}}
\small
        \caption{Overview of the solar flare events. Red (blue) dots mark the eruptive (confined) events, which are bracketed by the $45^\circ$ range from the central meridian. A list of the events and corresponding magnetic parameters is available online.}
\label{fig-overview}
\end{figure}

In previous studies, researchers have obtained a large number of eruptive and confined solar flares by checking the soft X-ray (SXR) flare catalog recorded by the Geostationary Operational Environmental Satellite (GOES) system \cite[]{2017ApJ...845...49K, 2023ApJ...958..104K, WangD2017ApJ...843L...9W, Toriumi2017ApJ...834...56T, Jing2018ApJ...864..138J}. The association of the flares with CMEs was determined by collating the LASCO CME catalog \cite[]{2009EM&P..104..295G} of the Solar and Heliospheric Observatory (SOHO)/Large Angle and Spectrometric Coronagraph (LASCO). The identification of a flare as being eruptive or confined depends on the alignment between the flare’s occurrence and the detection time of the CME, as well as the spatial correspondence between the flare's location and that of the CME.For our study, we collected 277 flares of GOES-class M1.0 and above  (a list of the events is available online); 267 of them is compiled from the above-mentioned studies covering flares from 2010 to 2019, whose flare types have been determined by the authors and further reconfirmed by us, and 10 additional events after 2019 are newly identified by us. All the flare events occur within $45^{\circ}$ from the central meridian, so that the measured magnetic field is reliable. Of the 277 events, 135 are confined ($\sim48.7\%$) and 142 are eruptive ($\sim51.3\%$). Figure \ref{fig-overview} shows an overview of the location of all the flare events, where eruptive events are marked by red dots and confined events by blue. The dashed lines indicate the $45^\circ$ longitudinal range from the central meridian. 

For each event in the dataset, we use a pre-flare vector magnetogram from Space Weather Helioseismic and Magnetic Imager AR Patches \cite[SHARP;][]{2014SoPh..289.3549B} to calculate the physical properties. The magnetic field vectors are presented as ($B_{r},B_{\theta},B_{\phi}$) in  heliocentric spherical coordinates in SHARP product, and the components are approximated to ($B_{x}, -B_{y}, B_{z}$) in heliographic Cartesian coordinates for subsequent calculation \cite[]{1990SoPh..126...21G, 2013arXiv1309.2392S}. 

\subsection{Calculation of the magnetic parameters}
\label{SS-para}

For all the 277 flare events in our dataset, we calculated various magnetic parameters by using the SHARP data, in which only pixels with $|B_{z}|>150~\mathrm{G}$ are considered in this study. The region of the AR before each flare is determined by a quadrilateral bounding the AR. The average magnetic field strength ($B_\mathrm{m}$) of the AR is the mean value of the magnetic field strength across all pixels in the AR. The total unsigned magnetic flux ($\Phi$) is computed by the product of the absolute value of $B_z$ and the area of the AR. According to the paper by \cite{WangD2017ApJ...843L...9W}, the half distance between the centroids of the positive and negative magnetic fields in the AR shows a strong correlation with the critical height of the torus instability above the PIL. This parameter ($d$) is calculated in our study for each AR. We also calculated the average magnetic shear angle ($\Theta$) along the polarity-inversion lines (PILs), the total photospheric free magnetic energy ($E_\mathrm{f}$), the free energy contained in the PIL region ($E_\mathrm{f_{PIL}}$), various parameters regarding HiFERs \cite[]{2020Sci...369..587K}, including the free energy contained in HiFERs ($E_\mathrm{f_{Hi}}$), the area of HiFERs ($A_\mathrm{Hi}$), and the area of the largest HiFER ($A_\mathrm{H1}$), as well as the ratio of the direct current and return current ($|\mathrm{DC/RC}|$) as an evaluation of the neutralization of electric currents within the AR. 

\begin{figure} [ht]
\centerline{\includegraphics[width=1\textwidth,clip=]{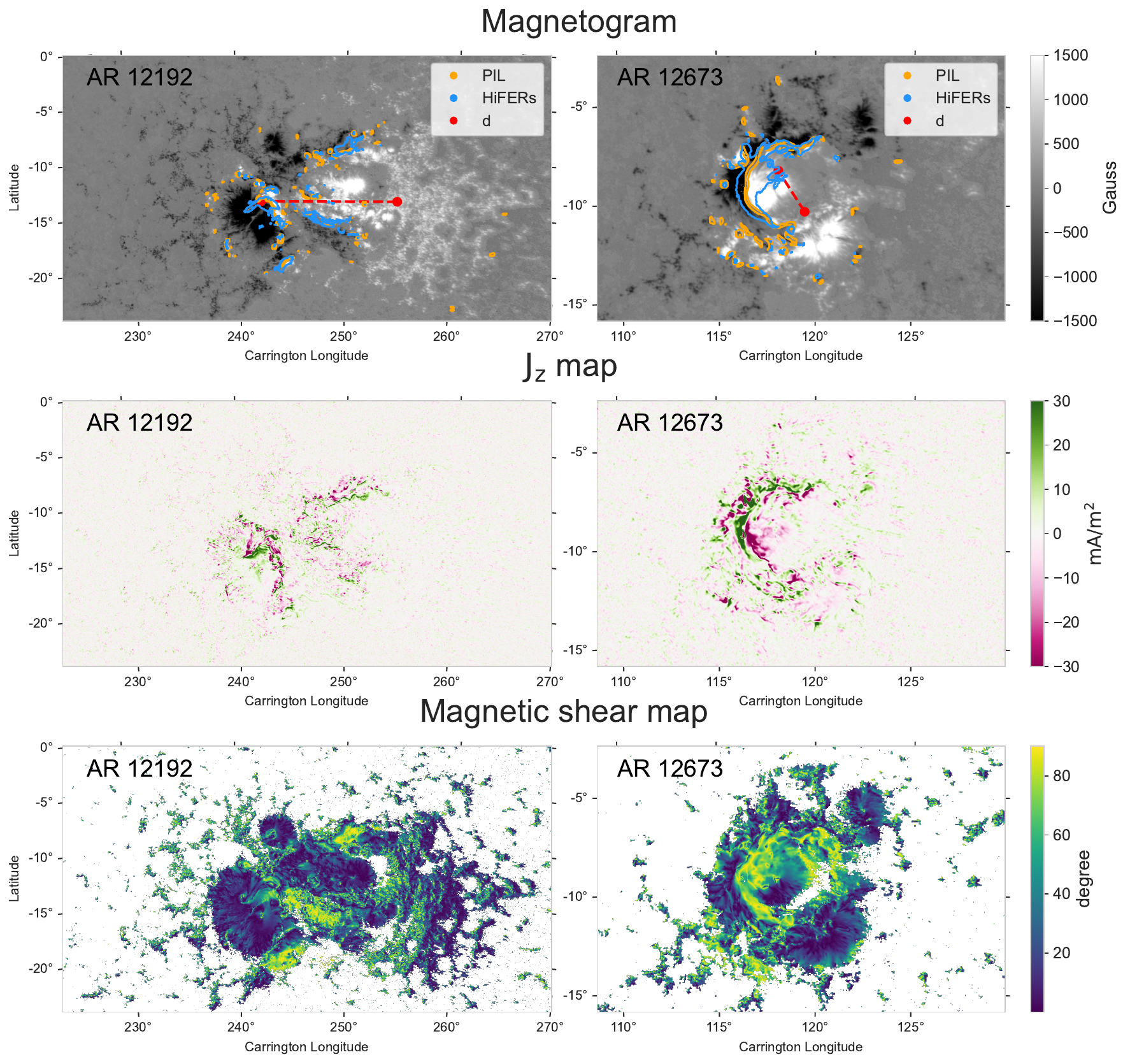}}
\small
        \caption{Example of the calculation results of all the parameters in AR 12192 (first column) and AR 12673 (second column). The top row shows the magnetograms of ARs. Red dots and dashed lines indicate the flux weighted centers and centroid distance. Orange color indicates the PIL regions and blue color outlines the HiFERs. The middle row shows the maps of $J_Z$. Magenta color indicates the negative $J_z$ while green color represents positive $J_z$. The last row shows the maps of the magnetic shear. The regions where the absolute value of magnetic field weaker than 150 $\mathrm{G}$ are shown in white.}
\label{fig-example} 
\end{figure}

In order to calculate the magnetic shear angle along the PILs, we first have to locate the PIL region in each magnetogram. We mainly followed the method described by \cite{2023ApJS..265...28J} and \cite{2020ApJ...893..123A}. First, we select all the  pixels with $|B_z|$ larger than 150 G, and then remove the regions of size smaller than 14 pixels to eliminate the noise-like regions. Next, a Canny edge operator \cite[]{4767851} is used to detect the boundaries between the positive and negative polarities. After dilating the intersection of the opposite polarities by a $5 \times 5$ kernel, the mask of PIL is finally acquired. An example of this mask is shown in orange in Figure \ref{fig-example}.

The magnetic shear angle is defined as the angular difference between the observed magnetic field ($B_{o}$) and the potential field ($B_{p}$) \cite[]{1984SoPh...91..115H, 1994ApJ...424..436W, 2007ApJ...662L..35Z}. The potential field is computed from the observed vertical component of the observed magnetic field using the Green function method \cite[]{1977ApJ...212..873C}. We only consider the pixels with $|B_z|>150$~G. Then, the magnetic shear angle ($\Theta$) is calculated following previous studies \cite[]{1984SoPh...91..115H, 1994ApJ...424..436W}:
\begin{equation}
    \Theta = \arccos \left( \frac{\mathbf{B_{o}} \cdot \mathbf{B_{p}}}{|B_{o}B_{p}|} \right).
\end{equation} 

The energy released during a flare comes from the non-potential magnetic field $B_{np}=|\mathbf{B_{o}}-\mathbf{B_{p}}|$ \cite[]{2020Sci...369..587K}. It was found that the free energy contained in an AR is highly correlated with the magnitude of the upcoming flares \cite[]{2014ApJ...788..150S, 2021ApJ...917L..29L}. Here, we calculated the total photospheric free magnetic energy by the integration of $B_{np}$ over the whole AR as follows \cite[]{ChenAQ2012, 2021ApJ...917L..29L}, 
\begin{equation}
    E_{f}=\frac{1}{8\pi}\int B_{np}^{2}\,dA.
\end{equation} 

\cite{2020Sci...369..587K} defined the regions in the AR where $ B_{np} > 1000$~G as high free energy regions (HiFERs) and deemed HiFERs as regions where flares likely occur. The HiFERs usually form islands that are distributed intermittently within an AR, and \cite{2020Sci...369..587K} ranked them by their area so that the first HiFER is the largest in each AR. Here we use the area of all HiFERs ($A_{Hi}$) and the area of the first HiFER ($A_{H1}$) as one of the parameters of non-potentiality of the ARs. Figure \ref{fig-example} shows HiFERs in two representative ARs. With the mask of PILs and HiFERs, we computed $E_{f_\mathrm{PIL}}$ and $E_\mathrm{f_{Hi}}$ for all the events.

In order to measure the non-neutralization of electric current for each AR, we first calculate the vertical electric current density 
\begin{equation}
    J_z=\frac{1}{\mu_0} (\frac{\partial B_y}{\partial x}-\frac{\partial B_x}{\partial y}).
\end{equation}
Then, the direct current (DC) and return current (RC) are computed in each polarity by integrating the $J_z$ values. We follow \cite{2014ApJ...782L..10T} to determine the sign of $J_z$ associated with DC and RC: $J_z$ of the dominant sign in the vicinity of PIL in each polarity is taken as DC. After the DC and RC in each polarity are computed, the non-neutralization of electric current is evaluated by
\begin{equation}
    |\mathrm{DC/RC}|=\frac{|\mathrm{DC}|^{+}+|\mathrm{DC}|^{-}}{|\mathrm{RC}|^{+}+|\mathrm{RC}|^{-}},
\end{equation}
where the superscripts `+' and `-' indicate the values in positive polarity and negative polarity, respectively \cite[]{2017ApJ...846L...6L, 2020ApJ...893..123A}. Table \ref{tbl-parameter} lists the description and sources of all the parameters mentioned above. Although these parameters have been investigated separately in previous studies, their relative relevance to the two flare categories of interest has not been evaluated though rigorous statistical tests. With this large dataset containing exclusively strong flares of GOES-class M1.0 and above, we expect to obtain statistically robust results.

\begin{deluxetable}{>{\centering\arraybackslash}p{0.15\textwidth} >{\centering\arraybackslash}p{0.35 \textwidth} >{\centering\arraybackslash}p{0.15\textwidth} >{\centering\arraybackslash}p{0.3\textwidth}}[ht]

\tabletypesize{\normalsize}
\tablewidth{\textwidth}

\tablecaption{List of the magnetic parameters.\label{tbl-parameter}}

\tablehead{
\colhead{Variable} & \colhead{Description} & \colhead{Unit} & \colhead{Source}
 }

\startdata 
$d$ & Half of the centroid distance of AR & Mm & \cite{WangD2017ApJ...843L...9W}\\
$\Phi$ & Total unsigned magnetic flux & Mx & ...\\
$B_\mathrm{m}$ & Mean magnetic field strength & Gauss & ...\\
$\Theta$ & Mean magnetic shear along PIL & degree & \cite{1984SoPh...91..115H}\\
$E_\mathrm{f}$ & Total free magnetic energy & $ \mathrm{erg}\cdot \mathrm{cm}^{-1}$ & \cite{ChenAQ2012} \\
$E_\mathrm{f_{PIL}}$ & Free magnetic energy within PIL  & $ \mathrm{erg}\cdot \mathrm{cm}^{-1}$ & \cite{ChenAQ2012}\\
$E_\mathrm{f_{Hi}}$ & Free magnetic energy within HiFERs  & $ \mathrm{erg}\cdot \mathrm{cm}^{-1}$ & \cite{2020Sci...369..587K}\\
$A_\mathrm{Hi}$ & Area of all the HiFERs & cm$^2$ & \cite{2020Sci...369..587K}\\
$A_\mathrm{H1}$ & Area of the H1FER & cm$^2$ & \cite{2020Sci...369..587K}\\
$|\mathrm{DC/RC}|$ & Non-neutralization of current & ... & \cite{2017ApJ...846L...6L}
\enddata
\end{deluxetable}

\section{Statistical Results}\label{S-sre}      

\subsection{Distributions of the parameters for all the events}
\label{SS-distribution_all}

\begin{figure} [ht]
\centerline{\includegraphics[width=1\textwidth,clip=]{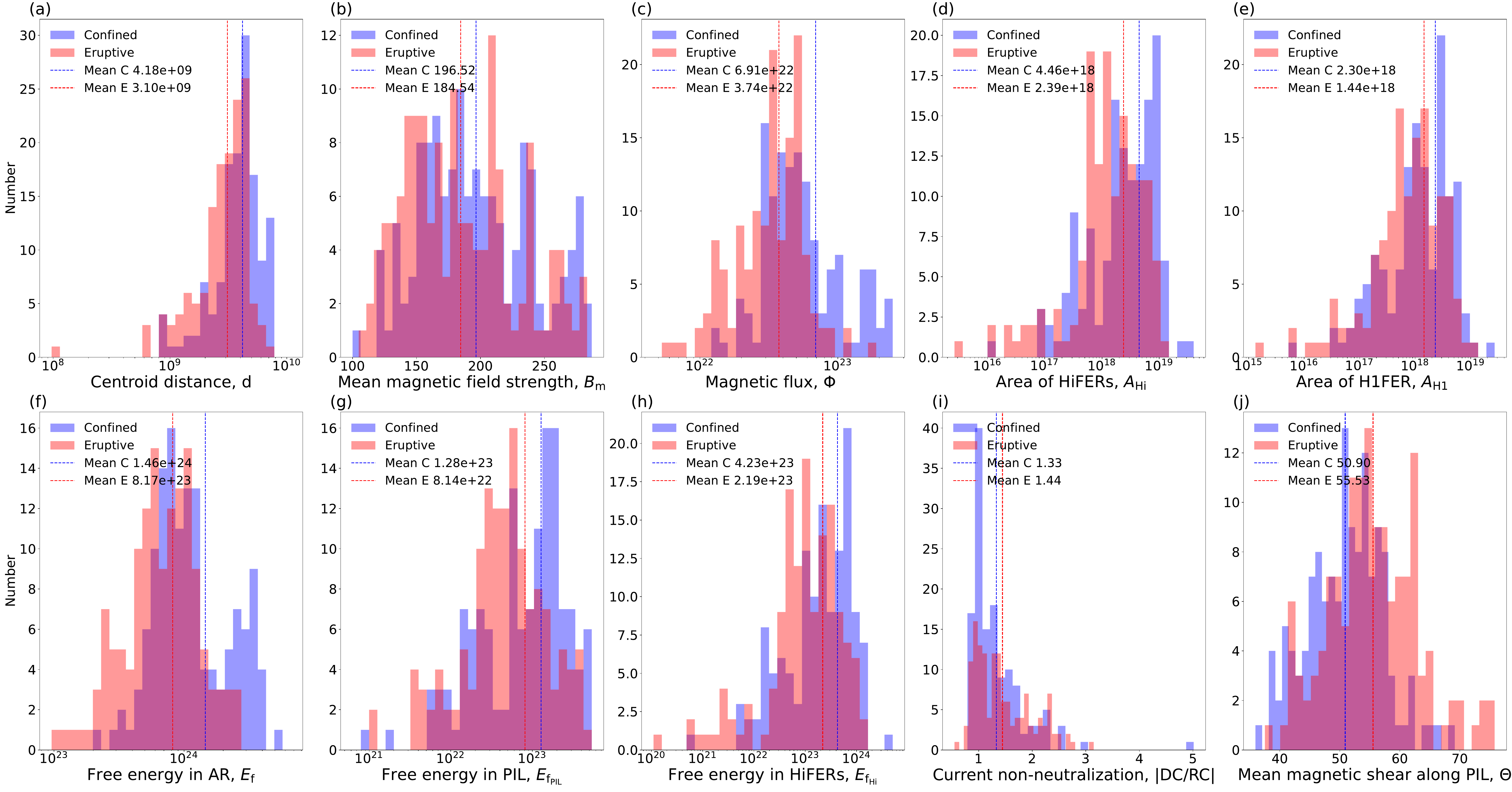}}
\small
        \caption{Distribution of the physical parameters for all the events. The red bars represent eruptive flares and the blue bars represent the confined flares. The overlapping portion is shown in deep brownish red. The dotted lines indicate the mean values of the parameters for eruptive flares (red) and confined flares (blue).}
\label{fig-distribution_all}
\end{figure}

Figure \ref{fig-distribution_all} shows the histograms of all the physical parameters mentioned above for all the events. The red bars represent the eruptive flare events while the blue bars the confined ones, and their mean values are indicated by the vertical dashed lines in the red and blue colors, respectively. We first analyze the magnetic parameters of the overall active region, namely the centroid distance, mean magnetic field strength, and total unsigned magnetic flux. We observe that the ARs hosting confined flares tend to have a longer centroid distance (Figure \ref{fig-distribution_all}a), a higher value of mean magnetic field strength (Figure~\ref{fig-distribution_all}b), and a larger value of total unsigned magnetic flux within the AR (Figure~\ref{fig-distribution_all}c). As discussed by \cite{WangD2017ApJ...843L...9W}, the half centroid distance of an AR is highly correlated with the critical height in torus instability \cite[]{2006PhRvL..96y5002K}. With the increasing centroid distance, the critical height tends to be higher, indicating that a magnetic flux rope has to rise to a higher altitude for the torus instability to set in, thus favoring the confinement of a flare \cite[]{2007ApJ...665.1428W, Liu2008ApJ...679L.151L}. The differences in centroid distance, magnetic field strength, and total unsigned magnetic flux between these two flare types align with the magnetic cage scenario. \cite{2018Natur.554..211A} performed an MHD simulation and showed that a weaker magnetic cage would produce a more energetic eruption associated with a CME. 

Next, we focus on the distributions of magnetic parameters around the PILs of these two flare types (Figure~\ref{fig-distribution_all}(d-j)). Figure~\ref{fig-distribution_all} (d-h) reveal notable differences between eruptive and confined flares in the distributions of the area of HiFERs, the area of the largest HiFER, magnetic free energy contained in HiFERs, magnetic free energy in PIL regions, and magnetic free energy in the entire AR. The mean values of these parameters show that the ARs hosting confined flares tend to have a larger value of magnetic free energy in the PIL regions, in the HiFERs, as well as in the whole AR. 
Figure~\ref{fig-distribution_all} (i) and (j) display the distributions of current non-neutralization and of magnetic shear along the PIL, respectively. In terms of the sample mean, ARs hosting eruptive flares exhibit both larger values of current non-neutralization and of magnetic shear along PIL than those hosting confined flares (Figure~\ref{fig-distribution_all}(i)). However, the difference between the two flare types are indistinct in current non-neutralization, with both being closely distributed around unity. On the other hand, the difference in magnetic shear along the PIL seems significant, which indicates stronger non-potentiality in the core flux of ARs hosting eruptive flares than those hosting confined ones. 

\subsection{Statistical Test}
\label{SS-test}

For a quantitative comparison between the eruptive and confined flares, we performed a statistical test in this section. To assess whether the distribution of each parameter follows a normal distribution so that a parametric statistical test is applicable, we performed the Kolmogorov–Smirnov test. The results indicated that only the distributions of PIL magnetic shear, mean magnetic field strength, and centroid distance are statistically consistent with a normal distribution. To maintain consistency in the subsequent statistical analysis, we applied a nonparametric statistical test known as the Mann-Whitney U test.

The Mann-Whitney U-test is used to compare two independent samples. By ranking all the values from both groups together and then comparing the sum of ranks between the two groups, the test evaluates whether one sample tends to occupy systematically higher (or lower) ranks than the other. \cite[]{2024ApJ...973...50L}. In order to test whether there exist statistically significant differences between the eruptive and the confined flares in the parameters mentioned above, we define the null hypothesis as follows: there is no significant difference in the physical parameters of hosting ARs between the eruptive and confined flare events. The difference between the ranks of the two flare types is then quantified using two U statistics:
\begin{equation}
    U_i=n_1 n_2+\frac{n_i (n_i +1)}{2}-\sum R_i, \quad i=1,2,
\end{equation}
where $n_i$ denotes the number of each type of flare events, and $\sum R_i$ represents the sum of ranks for each sample. The Mann-Whitney U test statistic, which is the smaller one of the two U statistics, is used to calculate the z-score under the normal approximation:
\begin{equation}
    z=\frac{U_i -\overline{x_U}}{S_U},
\end{equation}
with the mean $\overline{x_U}=n_1 n_2/2$ and the standard deviation $S_U=\sqrt{n_1 n_2 (n_1 + n_2)/12}$. At a significance level of $\alpha = 0.05$, the null hypothesis cannot be rejected if the test statistic $z$ falls within the range of $-1.96 \leq z \leq 1.96$ for a two-tailed test. The Mann-Whitney U test results for the magnetic parameters of ARs hosting eruptive and confined flare are presented in Table \ref{tbl-all_data}. The z-scores for all parameters fall outside the critical range, indicating a statistically significant difference between the two flare samples. Specifically, these results suggest that the parameters are correlated with the type of solar flares, and we subsequently calculate the effect size (ES) to assess the strength of this correlation. The effect size (ES) is computed as:
\begin{equation}
    \mathrm{ES}=\frac{|z|}{\sqrt{n}},
\end{equation}
where $n$ is the sample size. Following the conventions defined by \cite{Cohen}, the effect size is categorized as small (0.1), medium (0.3), and large (0.5). The effect size for each parameter is presented in the last column of Table \ref{tbl-all_data}. The analysis reveals that total unsigned magnetic flux, centroid distance, free energy in ARs, and PIL magnetic shear exhibit medium correlations with flare type, while other parameters show small correlations. Since a larger effect size indicates a stronger association between the parameters and flare type, total unsigned magnetic flux exhibits the strongest correlation with flare type among all the parameters examined. 

\begin{figure} [ht]
\centerline{\includegraphics[width=1\textwidth,clip=]{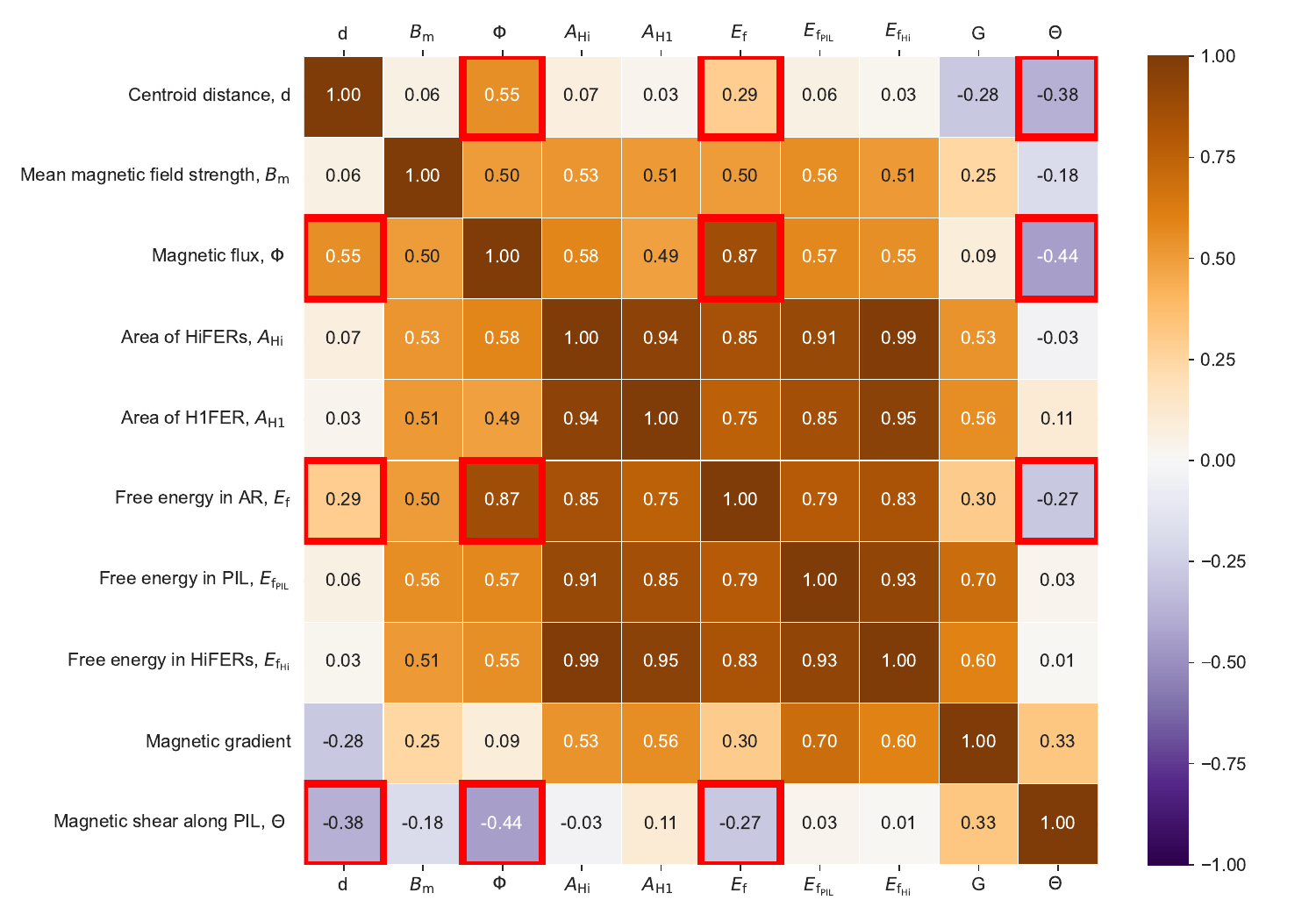}}
\small
        \caption{Correlation matrix showing Spearman correlation coefficient between different physical parameters for all the flares investigated. The correlation coefficients between $\Phi$,  $E_\mathrm{f}$, $d$, and $\Theta$ are highlighted by red bounding boxes.}
\label{fig-correlation_matrix} 
\end{figure}

To illustrate the relationship between the parameters of all the flares, the matrix in Figure~\ref{fig-correlation_matrix} shows the  Spearman correlation coefficients between the physical parameters in Table~\ref{tbl-parameter}. The correlation coefficients of $\Phi,\ E_\mathrm{f},\ \mathrm{d},\Theta$ are highlighted by red bounding boxes, as these parameters exhibit the strongest correlations with flare type among all parameters. The total unsigned magnetic flux shows a strong correlation with magnetic free energy in the AR ($r_s =0.87$), consistent with \cite{2021ApJ...917L..29L}. It is also positively correlated with the centroid distance  ($r_s=0.55$), while the remaining highlighted correlation coefficients are relatively weak ($r_s < 0.5$).

\begin{deluxetable}{>{\centering\arraybackslash}p{7cm}>{\centering\arraybackslash}p{3cm}>{\centering\arraybackslash}p{3cm}}[ht]

\tabletypesize{\normalsize}
\tablewidth{\textwidth}

\tablecaption{Mann-Whitney U test results for all the flare events.\label{tbl-all_data}}

\tablehead{
\colhead{Parameter} & \colhead{$z$-score\tablenotemark{a}} & \colhead{ES\tablenotemark{b}}
 }

\startdata 
Total unsigned magnetic flux, $\Phi$ [Mx] & -6.22 & 0.37 \\
Free energy in AR, $E_\mathrm{f}$ [erg/cm] & -5.55 & 0.33 \\
Centroid distance, $d$ [cm] & -5.32 & 0.32 \\
PIL Magnetic shear angle, $\Theta$ [deg] & 5.11 & 0.31 \\
Area of HiFERs, $A_\mathrm{Hi}$ [cm$^2$] & -3.93 & 0.24 \\
Free energy in HiFERs, $E_\mathrm{f_{Hi}}$ [erg/cm] & -3.58 & 0.22 \\
Free energy in PIL, $E_\mathrm{f_{PIL}}$[ erg/cm] & -3.56 & 0.21 \\
Area of 1st HiFER, $A_\mathrm{H1}$ [cm$^2$] & -2.85 & 0.17 \\
Mean magnetic field strength, $B_\mathrm{m}$ [G] & -2.24 & 0.13 \\
Current non-neutralization, $|\mathrm{DC/RC}|$ & 2.07 & 0.13 \\
\enddata

\tablenotetext{a}{$z$-score indicates how many standard deviations a data point is from the sample mean; given a confidence level of 95\%, the difference between the two flare types is insignificant if $|z|\le 1.96$}
\tablenotetext{b}{The effective size ES measures the degree of association between the two flare types; conventionally, small = 0.1, medium = 0.3, large = 0.5.}

\vspace{-0.5cm}
\end{deluxetable}

\subsection{Distributions of the parameters for events with a high-gradient PIL}
\label{SS-distribution_sub}
We select the flares that possess a PIL with strong magnetic gradient (termed `high-gradient PIL' hereafter) to compose a subset, and apply the same statistical test on this subset. This is motivated by previous studies showing that magnetic gradient along PILs is highly correlated with the productivity of solar flares \cite[]{2006ChJAA...6..477W, 2006SoPh..237...45C, ChenAQ2012, 2019ApJ...871...67C} and the productivity of CMEs \cite[]{2007A&A...462.1121G, Bobra&Ilonidis2016ApJ...821..127B}. 

\begin{figure} [ht]
\centerline{\includegraphics[width=1\textwidth,clip=]{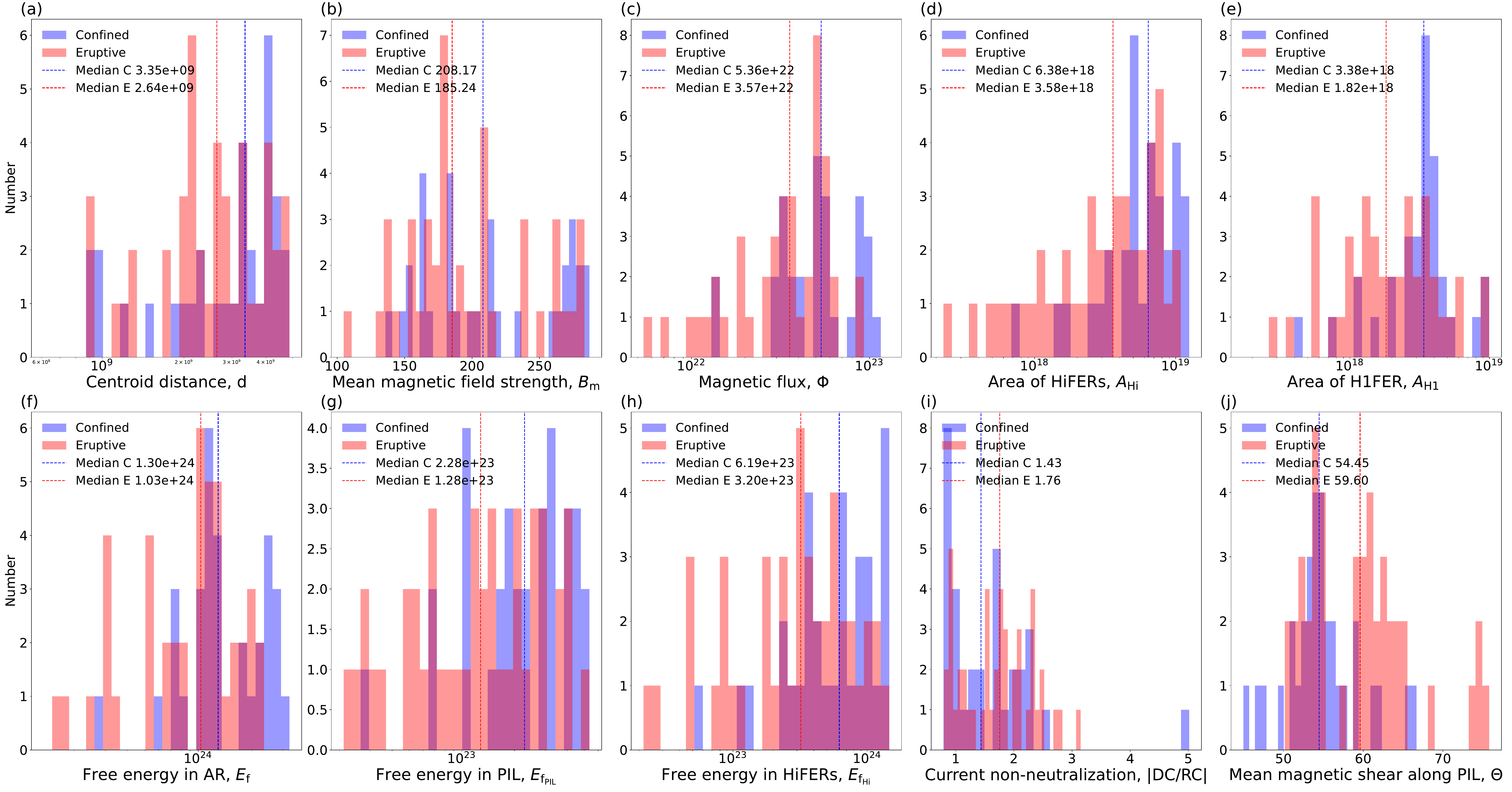}}
\small
        \caption{Distribution of the physical parameters for a subset of flares with high-gradient PILs. The red bars represent eruptive flares and the blue bars the confined flares. The overlapping portion is shown in deep brownish red. The dotted lines indicate the medians of the parameters for eruptive flares (red) and confined flares (blue).}
\label{fig-distribution_subset} 
\end{figure}

For each flare event, we calculated the average magnetic gradient along all the PILs in the AR \cite[]{2003JGRA..108.1380F}, which is reduced to a one-pixel width using a thinning algorithm \cite[]{10.1145/357994.358023}. 77 of the 277 flares, whose magnetic gradient along the PIL is above a threshold of 400 G~Mm$^{-1}$, are included in the subset of high magnetic gradient events. In the subset, there are 45 eruptive flares and 32 confined flares from 27 ARs. Of the two ARs shown in Figure \ref{fig-example}, flares originating from AR 12192 are excluded from this subset due to the relatively small magnetic gradient along the PIL, while those from AR 12673 are included. Furthermore, most of the ARs in this subset are similar to AR 12673, in that the flaring PIL is elongated and the flux concentration of opposite polarities are adjacent to each other. As a result, 8 super active regions (SARs) are included in the subset, namely, ARs 11158, 11302, 11429, 11515, 11967, 12205, 12297, and 12673 \cite[][]{ChenAQ2012, 2016ApJ...826..119L, 2019ApJ...871...67C, 2024ApJ...960...36D}; and they generated 67.5\% of the flares in the subset. Hence, it should be kept in mind that the statistical results of the subset are more or less biased toward these SARs. Figure~\ref{fig-distribution_subset} shows the distributions of the magnetic parameters for the events in this subset. However, the average magnetic gradient along all the PILs in the ARs shows no statistical significance in determining the eruptive behavior of flares ($|\text{z-score}|<1.96$).

From Figure~\ref{fig-distribution_subset}, one can see that the ARs hosting eruptive flares tend to exhibit a larger centroid distance, higher mean magnetic field strength, and greater current non-neutralization compared to those hosting confined flares. However, these differences appear less pronounced compared to those in the entire dataset, whereas other physical parameters seem to exhibit more pronounced differences between the two types of flares.

To assess how significant the differences in these parameters are between the two types of flare events, we performed the Mann-Whitney U test on this subset as well. The results of the Mann-Whitney U test are presented in Table \ref{tbl-subset}. Under the significance level $\alpha =0.05$, the critical range of z-score is [-1.96, 1.96]. From Table \ref{tbl-subset} one can see that, the z-scores of current non-neutralization, centroid distance and mean magnetic field strength all fall inside the critical range, which indicates that we cannot reject the null hypothesis that there is no significant difference between eruptive and confined flare events in terms of these three parameters. The z-scores of other parameters fall outside this critical range, and their corresponding ES values all exceed 0.3, suggesting a medium-level correlation with the flare types. Among the parameters, it is the area of HiFERs that has the largest ES value and is hence most strongly associated with the eruptive behavior of solar flares in this subset.

\begin{deluxetable}{>{\centering\arraybackslash}p{7cm}>{\centering\arraybackslash}p{3cm}>{\centering\arraybackslash}p{3cm}}[ht]

\tabletypesize{\normalsize}
\tablewidth{\textwidth}

\tablecaption{Mann-Whitney U test results for a subset of flares with high-gradient PILs.\label{tbl-subset}}

\tablehead{
\colhead{Parameter} & \colhead{z-score} & \colhead{ES}
 }

\startdata 
Area of HiFERs, $A_\mathrm{Hi}$ [cm$^2$] & -3.35 & 0.38 \\
Total unsigned magnetic flux, $\Phi$ [Mx] & -3.15 & 0.36 \\
Area of H1FER, $A_\mathrm{H1}$ [cm$^2$] & -3.03 & 0.35 \\
Free energy in AR, $E_\mathrm{f}$ [erg/cm] & -3.00 & 0.34 \\
Free energy in PIL, $E_\mathrm{f_{PIL}}$ [erg/cm] & -3.00 & 0.34 \\
Free energy in HiFERs, $E_\mathrm{f_{Hi}}$ [erg/cm] & -2.99 & 0.34 \\
PIL Magnetic shear [deg] & 2.85 & 0.33 \\
Current non-neutralization, $|\mathrm{DC/RC}|$ & 1.68 & ... \\
Centroid distance, $d$ [cm] & -1.44 & ... \\
Mean magnetic field strength, $B_\mathrm{m}$ [G] & -1.41 & ... \\
\enddata
\tablenotetext{a}{$z$-score indicates how many standard deviations a data point is from the sample mean; given a confidence level of 95\%, the difference between the two flare types is insignificant if $|z|\le 1.96$}
\tablenotetext{b}{The effective size ES measures the degree of association between the two flare types; conventionally, small = 0.1, medium = 0.3, large = 0.5.}
\vspace{-0.5cm}
\end{deluxetable}

\section{Discussion and Conclusion}\label{S-discussion}

To summarize, we have conducted a statistical analysis of the photospheric magnetic parameters of ARs prior to imminent major flares. Our statistical results show that ARs hosting eruptive flares tend to have smaller total unsigned flux ($\Phi$), total free magnetic energy ($E_\mathrm{f}$), flux-weighted centroid distance ($d$), but stronger magnetic shear along the PIL ($\Theta$) than those hosting confined flares (Sections \ref{SS-distribution_all}–\ref{SS-test}). These findings naturally fit into a “core–cage” framework, where $\Phi$ and $d$ mainly reflect the confinement of the magnetic cage, with $\Theta$ indicating the non-potentiality of the core field. The total unsigned magnetic flux shows the strongest correlation with the eruptive behavior of flares (Table~\ref{tbl-all_data}). An AR with a larger total unsigned magnetic flux not only stores more magnetic free energy ($r_s =0.87$; Fig.~\ref{fig-correlation_matrix}), but is associated with larger centroid distance ($r_s =0.55$) and smaller magnetic shear angle at the PIL ($r_s=-0.44$). As the half centroid distance is proportional to the critical height of the torus-instability \cite[]{WangD2017ApJ...843L...9W}, the larger centroid distance in confined flares implies a stronger constraining field, while the smaller PIL shear angle suggests that the magnetic free energy in such an AR is more likely attributed to the magnetic cage than to the core structure, as indicated by the negative correlation between the AR free energy and PIL shear angle ($r_s=-0.27$) as well as that between the centroid distance and PIL shear angle ($r_s=-0.38$). \cite{2021ApJ...917L..29L} also found that the AR-wide magnetic free energy is smaller for eruptive flares than for confined flares, which is interpreted as a manifestation of the constraining effect of the magnetic cage. Thus, ARs with a larger unsigned magnetic flux $\Phi$, which tend to possess a stronger magnetic cage and to accumulate a larger amount of magnetic free energy, are more likely to host flares without a subsequent CME. An extreme case in this regard is AR 12192, whose huge amount of magnetic free energy contributes to numerous flares but no CME during its disk passage \cite[]{Sun2015ApJ...804L..28S,2016ApJ...826..119L}. 


Regarding the current non‑neutralization, it differs between two types of flares statistically, when evaluated over the entire AR, but the correlation is weak (ES = 0.13). This agrees with the findings of \cite{2020ApJ...893..123A}, in which $|\mathrm{DC/RC|}$ likewise clusters close to 1 but differences still exist between flaring/CME-active ARs and flare/CME-quiet ARs. For completeness, we have also performed a similar analysis within flare ribbon masks, and demonstrated that eruptive flares show larger ribbon-masked $E_\mathrm{f}$ and $\Phi$ than confined flares, which is consistent with \cite{2023ApJ...958..104K}, but the differences of these two parameters between eruptive and confined flares are not statistically significant (both $z$-scores fall into the critical range). 


In the subset of 77 events whose mean PIL magnetic gradient exceeds 400 G~Mm$^{-1}$ (\S\ref{SS-distribution_sub}), most of the non‑potential parameters show a medium correlation with the flare type, whereas centroid distance, mean field strength, and current non‑neutralization lose statistical significance (Table~\ref{tbl-subset}). For these events, the median half–centroid distance ($2.75\times10^{9}$~cm), median magnetic‐field strength ($191.7$~G), and current non-neutralization (1.65) are significantly different from the corresponding values for the full dataset ($3.55\times10^{9}$~cm, $184.7$~G, and 1.25, respectively). These results indicate that the ARs in the subset are more compact, possess stronger magnetic fields, and have higher degree of current non-neutralization. Within this subset, the Mann–Whitney U test indicates that the area of HiFERs ($A_{\mathrm{Hi}}$) is the parameter most strongly tied to the flare type, i.e., confined flares tend to have larger $A_{\mathrm{Hi}}$ than eruptive ones, consistent however with the trend in the full dataset. In other words, if high free energy regions in the hosting AR spread more widely, a flare is more likely confined. 

It is noteworthy that our high magnetic gradient subset contains eight well-studied flare- and CME–productive ARs \cite[11158, 11302, 11429, 11515, 11967, 12205, 12297, and 12673;][]{2015SCPMA..58.5682W, 2016ApJ...826..119L, Toriumi2017ApJ...834...56T, 2018JSWSC...8A...9S, 2024ApJ...964..159L, 2024A&A...686A.115G}, while it excludes the famous flare-rich but CME-poor AR 12192 \cite[]{Sun2015ApJ...804L..28S,2016ApJ...826..119L}. This contrast suggests that both the average magnetic gradient along the PIL and the area of HiFERs are closely linked to AR 12192’s puzzling behavior.  On the other hand, these eight ARs generate the majority (67.5\%) of the flares in the subset, indicating a statistical bias toward SARs.

In conclusion, over the full sample, a four-parameter pre-flare combination, i.e., the unsigned magnetic flux ($\Phi$), magnetic free energy ($E_\mathrm{f}$), and centroid distance ($d$) of an AR, and the mean magnetic shear along the PIL in the AR ($\Theta$), yields the strongest discrimination between eruptive and confined flares; among the four parameters, the total unsigned magnetic flux $\Phi$ is most strongly correlated with the flare types. But for the subset with high-gradient PILs, the area of HiFERs ($A_{\mathrm{Hi}}$) becomes the most effective parameter in distinguishing confined from eruptive flares. These parameters may be promising in the future physics-based CME forecasting models. 

\begin{acknowledgments}
This work was supported by the Strategic Priority Program of the Chinese Academy of Sciences (XDB0560102), the National Key R\&D Program of China (2022YFF0503002), and the NSFC (42274204, 12373064, 42188101, 11925302). The authors thank the Joint Science Operations Center (JSOC) at Stanford University for the Space-weather HMI Active Region Patches (SHARP) data used in this study.
\end{acknowledgments}

\bibliographystyle{aasjournal}
\bibliography{sample7}



\end{document}